# The two-way knowledge interaction interface between humans and neural networks

Zhanliang He
*Xidian University*
Xi'an, China
22031212282@stu.xidian.edu.cn

Nuoye Xiong
*Xidian University*
Xi'an, China
xiongnuoye@gmail.com

Hongsheng Li
*Xidian University*
Xi'an, China
hsli@stu.xidian.edu.cn

Peiyi Shen
*Xidian University*
Xi'an, China
pyshen@xidian.edu.cn

Guangming Zhu
*Xidian University*
Xi'an, China
gmzhu@xidian.edu.cn

Liang Zhang
*Xidian University*
Xi'an, China
liangzhang@xidian.edu.cn

*Abstract*—Despite neural networks (NN) have been widely applied in various fields and generally outperforms humans, they still lack interpretability to a certain extent, and humans are unable to intuitively understand the decision logic of NN. This also hinders the knowledge interaction between humans and NN, preventing humans from getting involved to give direct guidance when NN's decisions go wrong. While recent research in explainable AI has achieved interpretability of NN from various perspectives, it has not yet provided effective methods for knowledge exchange between humans and NN. To address this problem, we constructed a two-way interaction interface that uses structured representations of visual concepts and their relationships as the "language" for knowledge exchange between humans and NN. Specifically, NN provide intuitive reasoning explanations to humans based on the class-specific structural concepts graph (C-SCG). On the other hand, humans can modify the biases present in the C-SCG through their prior knowledge and reasoning ability, and thus provide direct knowledge guidance to NN through this interface. Through experimental validation, based on this interaction interface, NN can provide humans with easily understandable explanations of the reasoning process. Furthermore, human involvement and prior knowledge can directly and effectively contribute to enhancing the performance of NN.

## I. INTRODUCTION

In recent years, neural networks (NN) have become increasingly common in research and applications, playing an important role in many fields such as medicine [1], security [2] and finance [3]. Given the breadth and importance of machine learning applications, it is crucial to ensure the interpretability and transparency of NN and the reliable use of machine learning models by humans. However, knowledge exchange between humans and NN becomes difficult due to the lack of effective interaction interfaces: humans cannot easily understand the internal reasoning logic of NN, and more importantly, when NN's decision-making is unreasonable, humans cannot participate in it to directly give guidance and affect the performance of NN. Therefore, the interpretability of NN and the exchange of knowledge between humans and NN are still two key issues that exist in machine learning models at present.

In terms of interpretability studies, gradient-based pixel-level interpretation methods [4], [5] can give decision explanations only at low-level perspective of pixels. Visual reasoning explanations (VRX) [6] provides human-readily interpretable visual explanations by simulating the reasoning process of the original NN, but still fail to produce two-way knowledge exchange and interaction with humans. In the field of human-neural networks interaction research, Human-In-the-Loop ML (HILML) [7] and Interactive Model Learning (IML) [8] focus on data by using manually managed data to train models, rather than interacting with and correcting the model's reasoning logic.

In order to solve these problems, we constructed a two-way knowledge interaction interface between humans and NN. As shown in Fig. 1, it mainly includes the following processes: (1) By using high-level visual concepts and their structural relationships to construct a graph as an abstract representation for each class, the class-specific structural concepts graph (C-SCG) will be used as a "language" for knowledge exchange between humans and NN. (2) Through NN-to-Human pathway, each image is similarly represented as an image-level structural concepts graph (I-SCG), and a graph network is constructed to learn and observe the contribution of all elements in I-SCG, in order to provide an intuitive explanation of NN's decision-making at a high level of conceptual perspective. (3) Through Human-to-NN path, humans can easily understand this explanation provided by NN and capture possible irrationalities in it, then directly modify this abstract C-SCG for each class. Afterwards, human's modified logic can be transferred to NN through methods such as knowledge distillation, thus intervening and improving the performance of NN.

## II. RELATED WORK

Interpretability: For some critical fields such as medical diagnostics [1] and autonomous driving [2], the interpretability of machine learning models is crucial, as rationality and credibility of decisions are essential elements. To overcome the interpretability challenges of NN, researchers have proposed various approaches. CAM [9] and Grad-CAM [4] use gradient correlation techniques to visualize which parts of the inputs are more sensitive when making decisions, thus helping to explain model's decision basis. VRX [6] explains the reasoning logic of NN with the help of abstract visual concepts. Based on this, in our work, we use that interpretation as a basis for two-way interaction between humans and NN.

Human-NN Interaction: Human prior knowledge and reasoning logic can refine and optimize NN's understanding in many scenarios. IML [8] centers on the machine learning model and lets humans act as a server to realize the interaction between humans and NN, but this will limit human involvement. "Tell me where to look" [10] corrects NN's segmentation errors based on interpretable attention maps, but this is still limited to a low level. In our work, we use high-level structural concepts graph (SCG) as the "language" to allow human users and NN to communicate with each other and exchange knowledge more effectively.

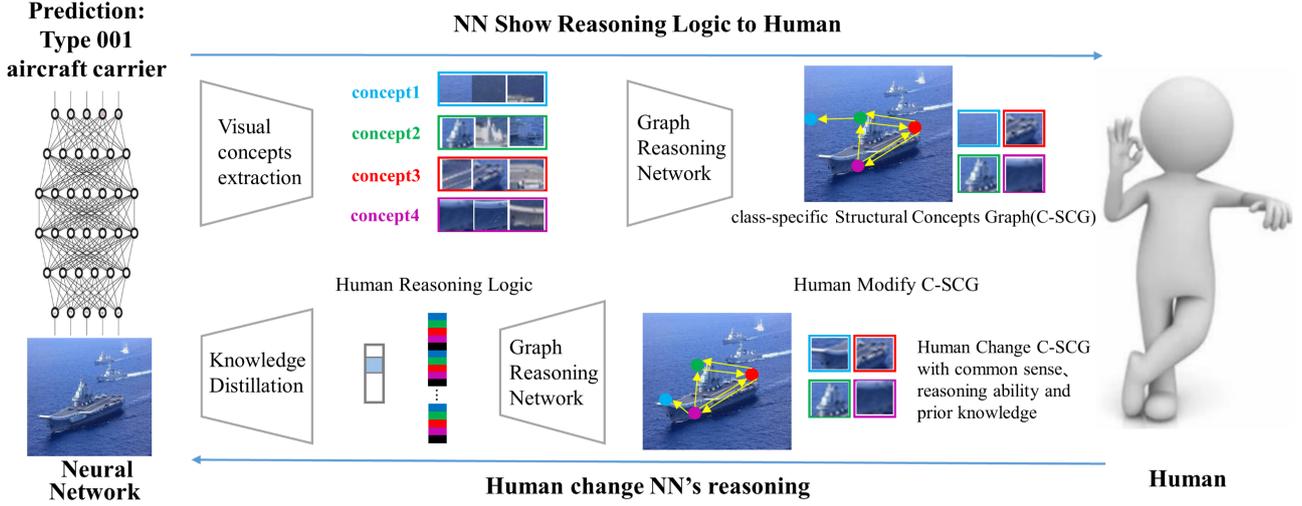

Fig. 1. Implementation of the two-way knowledge interaction interface between humans and neural networks.

Graph neural networks: Graph Neural Networks (GNN) are machine learning models used to process graph structured data [11]. And graph structured data consists of nodes and edges connecting the nodes, which can be used to represent complex semantic and structural information [12]. The core idea of GNN is to continuously aggregate information of nodes and their neighboring nodes, so as to extract global features of the structured graph step by step [13]. In our work, we construct a graph reasoning network (GRN) based on GNN to learn semantic relations between concepts and as a tool for two-way knowledge exchange.

Knowledge distillation: The purpose of knowledge distillation is to distill and extract the knowledge contained in a trained model into another model [14]. Due to its effectiveness and flexibility, knowledge distillation method has a wide range of applications in different fields such as model compression [15] and knowledge transfer [16]. In our work, we use knowledge distillation to transfer knowledge between GRN with human prior knowledge and NN.

## III. INTERACTION INTERFACE BETWEEN HUMAN AND NN

### A. NN provide Decision Explanations to Humans

NN provide humans with the reasoning logic for each decision. Specifically, NN first represents its understanding of each class as a specific SCG, where nodes represent multiple abstract concepts extracted by NN based on the class images and edges represent the structural relationships and dependencies between concepts. After that, we will use gradient-based method to calculate the contribution value of each element in I-SCG and explain the reasoning logic of NN based on this. The complete process is specifically shown in Fig. 2 and next we will explain each part in detail.

Visual concepts extraction: Inspired by the concept-based interpretation (ACE) technique [17], we use visual concepts to represent a given NN's perception of a specific class to help explain its underlying decision-making process. And we use a similar process to ACE when doing concepts extraction for specific classes: firstly, multiple images are selected for each class, and for each image, a multi-resolution super-pixel segmentation (e.g., SLIC [18]) is performed to obtain multiple patches images. Then these patches are converted into feature vectors in a specific convolutional layer of NN and clustered in the vector space. After that, each clustering result will be used as an abstract visual concept for that class and we will finally use a concept scoring method (e.g., TCAV [19]) to evaluate the importance of the concepts represented by each cluster. However, in this process, in order to improve the validity of the extracted concepts, we suggest constraining the extracted regions using a gradient attention method (e.g., Grad-CAM [4]) before performing segmentation, which constrains the relevant regions for concepts extraction to the foreground object portion and helps exclude irrelevant portions (e.g., most of the background). Based on the above process, we can use the original trained neural network to extract the top N visual concepts and their average feature vectors for each class of interest.

Construction of SCG: Based on the extracted visual concepts obtained for each class, we can construct a SCG to represent each class image. Specifically, given an image, patches are first obtained using the multi-resolution segmentation algorithm, after which the feature vectors of the patches in a particular convolutional layer are computed and compared with the average concept feature vectors of the K most important visual concepts of each class (e.g., K=4) filtered in the previous step. A patch will be recognized as a detected concept if the Euclidean distance between the patch features and the average concept features is less than a threshold value. Then, the K visual concepts obtained from matching are used as nodes of SCG and SCG is initially a fully connected graph $(V, E)$. Two attributes are assigned to each directed edge $(v_i, v_j) \in E$: one is the spatial relationship representation between the nodes, initialized with the spatial positions of the concept nodes at its two ends $[x_i, y_i, x_j, y_j]$, and the other one is the dependency $e_{ij}$ (a trainable value) between concepts $v_i$ and $v_j$. Such a design can help us make full use of valuable conceptual information and discover how visual concepts and their relationships affect the inference logic of NN. Also note that considering all n classes of interest, we need to have generated n SCGs for the same input image.

Constructing GRN to simulate the original NN: The graph neural network $g$ we constructed consists of two modules, the graph convolution $G$ used to learn graph representation of SCGs, and embedding network $E$ is used to fuse multi-class SCGs for final class prediction. In order to achieve decision interpretation, we need to ensure that GRN follows the same reasoning process as the original NN. Specifically, given an input image $I$ and a trained classifier $F$, as well as SCGs $h=\{h_1,h_1,...h_n\}$, obtained by matching the image for n classes respectively, we need to ensure the prediction consistency between GRN and NN, i.e.,

$$g(h)=E(G(h))=F(I) \quad (1)$$

so GRN then will be trained to imitate the original NN by minimizing Loss as:

$$Loss=||\sigma(G(h)-\sigma(F(I))||l_1 \quad (2)$$

where σ is a normalization function.

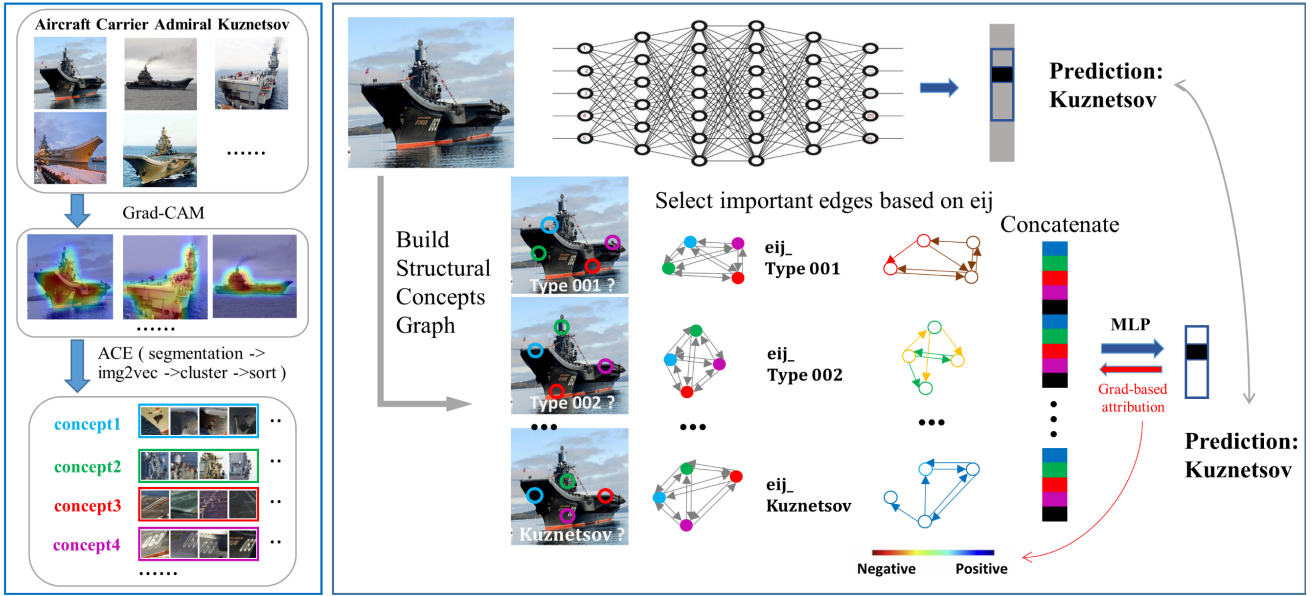

Fig. 2. NN provide decision explanations to human based on C-SCG of each class.

Explanation of decision-making: After that, we can use GRN to observe the flow of all elements in SCG and their contribution to the decision making, which will help us to realize decision-interpretation process of the original model. Specifically, for each input image $I$ and $h$, we can obtain the final prediction y in GRN. Then for each class c of interest, we have a class prediction score $y^c$ and compute the gradient of $y^c$ with respect to the graph embedding from the n hypotheses as:

$$grad_i = \frac{\partial p^c}{\partial G^i(h_i)}, \; i=1, 2, ...m \quad (3)$$

$grad_i$ represents the vector of contribution weights of hypotheses $h_i$, where $h_i$ is the SCG for class $i$. The contribution score $s_i$ of each hypothesis $h_i$ after that is the weighted sum of $grad_i$ and $G^i(h_i)$, as:

$$s_i=grad_i^T \; G^i(h_i), \; i=1,2,...m \quad (4)$$

based on the calculated contribution scores $s_i$, we can further represent the positive or negative contribution of each concept or the spatial relationship between them to the decisions made by the original NN.

### B. Humans provide Prior Knowledge to NN

After understanding the reasoning logic of NN, human users can judge its reasonableness based on prior knowledge and reasoning ability. If an error occurs, human users are able to actively correct decision logic by effectively updating the C-SCG (e.g., replacing a visual concept or changing the structural relationships between concepts). Once the updating process is complete, the modified C-SCG can be used to re-match concepts and generate graph information for each image, and then the new graph information will be used to train a GRN, which carries human knowledge and has a more reasonable perception of the "local logic" among the classes of interest. After that, we transfer the rational logic of this GRN to the original NN by knowledge distillation. The specific steps are as follows.

Human users modify C-SCG: The C-SCG represents NN's abstract cognition of each class, where each node represents a visual concept and each edge represents the structural relationship and dependency between two concepts. C-SCG can be modified intuitively to change NN's confusing perception of the "local logic" between similar classes. The modifications can occur at both nodes and edges, which correspond to replacing the visual concepts and changing the relationship between visual concepts, respectively: (1) concept modification: when the concept represented by a node is a background concept or is not unique (e.g., concepts such as wheels, which exist in many classes), other concepts can be selected for replacement from the pool of concept candidates consisting of all the concepts extracted during concepts extraction. (2) concept relationship modification: when the dependencies between concepts discovered by NN do not match human cognition, humans can directly delete these edges in C-SCG. Similarly, when the dependency relationship between two concepts is strong, humans can add an edge between these two visual concepts in C-SCG. And in practice, nodes and edges can be modified at the same time to handle more complex situations. And we only need to modify C-SCG, then use the modified C-SCG as a template to match training images, so that the modified logic can be applied to all image-level structural concepts graphs (I-SCGs).

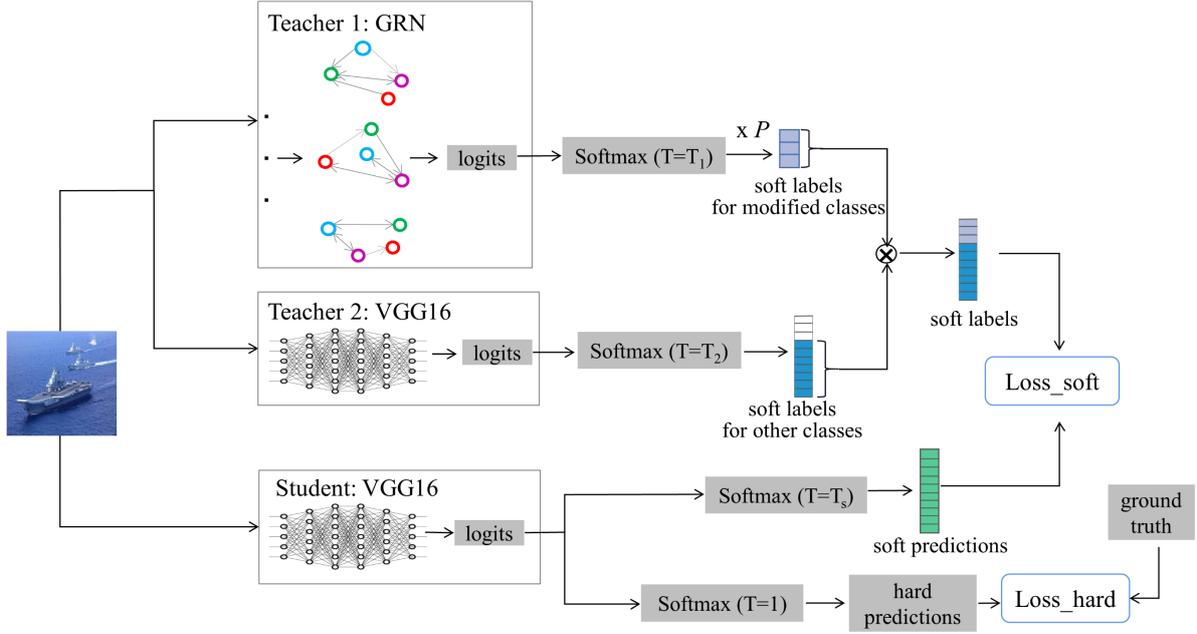

Fig. 3. The pipeline of knowledge distillation. GRN focuses on modified classes and the original NN focuses on other classes. And then we will use both soft labels and hard labels (ground truth) to train the student model.

Training GRN with modified I-SCGs: In general, the classes that need to be manually modified by human are some of the classes predicted by the original NN, for example, some similar classes are easy to confuse. And we usually only want to intervene in the structural concepts graphs of these classes to change NN's erroneous understanding of the "local logic" between them. Therefore, we will construct a GRN that only targets these classes of interest. So, after human modification process is complete, the process of GRN training is same as before: based on the modified C-SCGs, we can match all the training images to I-SCG for each modified class, and this graph information is already integrated with human user's knowledge. The I-SCGs then will pass through the graph convolutional backbone and MLP in GRN, which ultimately generates predictions and constructs cross-entropy loss between them and ground truth. After the training is completed, the final GRN will be equipped with human prior knowledge, which allows this GRN to generate more accurate classification decisions for class images based on the modified inference logic. In the later process, this GRN will act as a teacher network to transfer human knowledge to the original NN.

Knowledge distillation: The final step is to transfer the knowledge with human reasoning logic from GRN back to the original NN. Since GRN focuses only on our modified classes, we will use two teacher networks: GRN (teacher 1) focuses on modified classes and the original NN (teacher2) focuses on unmodified classes. In this way, we can improve the cognitive accuracy of modified classes without negatively affecting the categorization performance of other classes. Fig. 3 illustrates the specific implementation process of knowledge distillation: two teacher models together provide soft labels, GRN provides the probability of modified classes and the original NN with fixed parameters provides the probability of unmodified classes. The student network has the same architecture as the original NN and is initialized using the weights of it. Thus, the overall loss during knowledge distillation is as follows:

$$Loss = \alpha Loss_{soft} + \beta Loss_{hard} \quad (5)$$

where $\alpha$ and $\beta$ are the weighting of the two terms in distillation process. $Loss_{soft}$ is the first loss function, which represents the cross-entropy between combined soft labels provided by teacher models and probability distribution of the student model outputs, and:

$$Loss_{soft} = -\sum_{c}^{N} p_c^T log(q_c^{T_s}) \quad (6)$$

where $N$ denotes the total number of classes in the original NN, $p_c^T$ represents the probability value of class $c$ after the soft-labels combination of two teacher models, and $q_c^{T_s}$ is the probability value of class $c$ predicted by the student network at temperature $T_s$, and:

$$p_c^T = \begin{cases} p_c^{T_1} P, & (c \in S_p) \\ p_c^{T_2}, & (c \in S \setminus S_p) \end{cases} \quad (7)$$

$$p_c^{T_1} = \frac{exp(v_c^1/T_1)}{\sum_{k}^{N} exp(v_k^1/T_1)} \quad (8)$$

$$p_c^{T_2} = \frac{exp(v_c^2/T_2)}{\sum_{k}^{N} exp(v_k^2/T_2)} \quad (9)$$

$$q_c^{T_s} = \frac{exp(z_c/T_s)}{\sum_{k}^{N} exp(z_k/T_s)} \quad (10)$$

where $z_c$、$v_c^1$、$v_c^2$ denote the inputs of the student network, teacher1 and teacher 2, respectively; $T_1$ and $T_2$ represent the temperature values used in the knowledge distillation of two teacher models, respectively; $S$ denotes the set of the total classes in the original NN, and $S_p$ denotes the set of

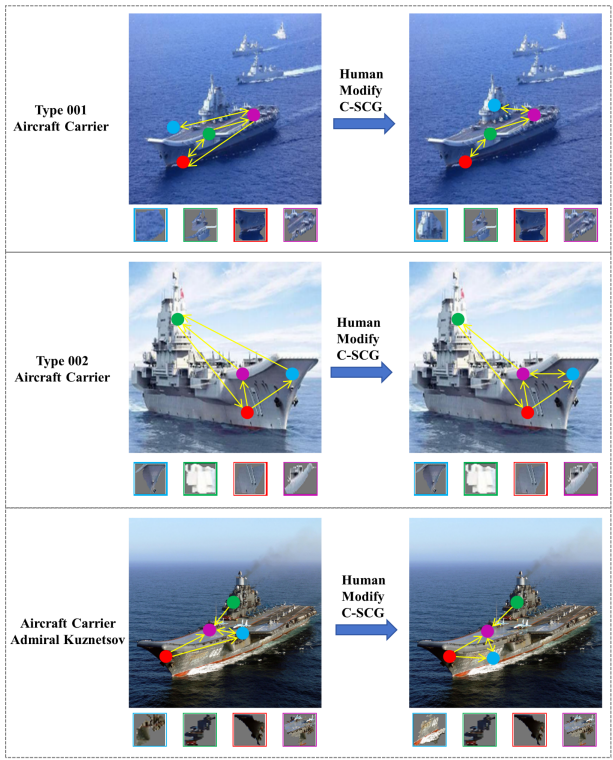

Fig. 4. Examples of human modification of C-SCGs. And we use one instance to visualize the states of C-SCGs before and after modification. (1) For the class "Type 001 aircraft carrier", we replace concept1 from a background concept to a "ship island" concept and removing the edge from the "deck" concept to the "ship's railing" concept. (2) For the class "Type 002 aircraft carrier", we remove the edge from the "cabin forward" concept (concept1) to the "ship island" concept (concept2) and adding a two-way edge between the "cabin forward" concept (concept1) and the "deck edge" concept (concept4). (3) For the class "Aircraft Carrier Admiral Kuznetsov", we replace concept1 from the "deck edge" concept to a more representative concept "hull number".

modification classes, $n=|S_p|$ represents the number of modification classes. And:

$$P = \sum p_c^{T_I}, (c \in S_p) \quad (11)$$

which denotes the probability proportion of the *n* modification classes in the original NN with respect to the total number of all classes *N*. $Loss_{hard}$ is the second loss function, represents the cross-entropy between the true class label of this image and the probability distribution of student model's output at *T=1*, as:

$$Loss_{hard} = -\sum_c^N g_c \, log(q_c^1) \quad (12)$$

where $g_c$ is the true class label of class *c*, $q_c^1$ is the probability value of class *c* in the student network prediction at *T=1*.

## IV. EXPERIMENTS

In order to validate the effectiveness of this two-way interaction interface, we constructed a dataset including a total of 30 classes of objects, covering aircraft carriers, submarines, tanks, fighter jets, helicopters and so on. Then we collected 70 images as a training set and 70 images as a test set for each class. After that, we trained a classification model using VGG-16 [20] as the original NN.

TABLE I. ACCURACY ON MODIFIED CLASSES OF ORIGINAL NN, BASELINE AND OUR METHOD

|  | *Type 001* | *Type 002* | *Admiral Kuznetsov* |
|---|---|---|---|
| Original NN | 0.69 | 0.71 | 0.59 |
| Baseline: **Without** human modification | 0.69 | 0.70 | 0.58 |
| Ours: **With** human modification | 0.74 | 0.76 | 0.66 |

By testing the class accuracy of this classification model, we found that our classification model has low accuracy in the classes of "Type 001 aircraft carrier", "Type 002 aircraft carrier", and "Aircraft Carrier Admiral Kuznetsov", which may due to a deviation in the model's understanding of the "local logic" between these three classes. So we will perform the above process for these three similar classes. We first visualize the inference logic of the original NN using the NN-to-Human path, based on which human users can directly check and modify C-SCG of each class. Fig. 4 shows the C-SCG comparison before and after human modification for these three classes. We then train the new GRN with human-updated C-SCGs. To transfer the human logic back to the original NN to improve its performance, we finally modify the neural network using knowledge distillation. Meanwhile, in order to assess to what extent human intervention improves the performance of the original NN, we performed knowledge distillation without human intervention by supplying unmodified C-SCGs directly to the Human-to-NN path as a baseline experiment.

The final test results are shown in the TABLE I. . For three modified classes, the accuracy of the modified student network in the test set is improved, the original NN's confusion for these three classes has been partially corrected. For the other 27 unmodified classes, the accuracy before and after distillation is nearly consistent (with an average accuracy of 89.3% before distillation and 89.7% after distillation). The results show that by merging human user inputs, our two-way knowledge interaction interface can accurately modify the reasoning logic of target classes while maintaining the reasoning logic for other classes.

## V. CONCLUSION

We constructed a two-way knowledge interaction interface between humans and neural networks (NN). NN provide intuitive logical explanations to humans based on abstract visual concepts and their structural relationships, which can help humans understand the internal reasoning of NN and discover possible deviations in them. After understanding the reasoning logic of NN, humans can directly improve and correct the "local logic" between similar classes and then transfer prior knowledge to NN through this interface. Our experiments show that the intervention of human reasoning ability can help correct NN's confusing understanding of "local logic" between some classes and provide direct guidance and assistance for its performance improvement. We believe that this provides new ideas and methods for the study of neural network interpretability and knowledge exchange between humans and neural networks.